\documentclass[aps,prl,superscriptaddress,amsmath,amssymb,twocolumn]{revtex4}
\usepackage{graphicx}
\usepackage{color,hyperref}

\definecolor{mygrey}{gray}{0.80}
\definecolor{darkblue}{rgb}{0.0,0.0,0.5}
\hypersetup{colorlinks,breaklinks,
            linkcolor=darkblue,urlcolor=darkblue,
            anchorcolor=darkblue,citecolor=darkblue}

\begin{document}

\title{Degenerate Quasicrystal of Hard Triangular Bipyramids}

\author{Amir Haji-Akbari}
\affiliation{Department of Chemical Engineering, University of Michigan, Ann Arbor, MI 48109}

\author{Michael Engel}
\affiliation{Department of Chemical Engineering, University of Michigan, Ann Arbor, MI 48109}

\author{Sharon C. Glotzer}
\email{sglotzer@umich.edu}
\affiliation{Department of Chemical Engineering, University of Michigan, Ann Arbor, MI 48109}
\affiliation{Department of Materials Science and Engineering, University of Michigan, Ann Arbor, MI 48109}

\date{\today}

\begin{abstract}
We report a degenerate quasicrystal in Monte Carlo simulations of hard triangular bipyramids each composed of two regular tetrahedra sharing a single face. The dodecagonal quasicrystal is similar to that 1	recently reported for hard tetrahedra [\href{http://www.nature.com/nature/journal/v462/n7274/full/nature08641.html}{Haji-Akbari et al., Nature (London) 462, 773 (2009)}] but degenerate in the pairing of tetrahedra, and self-assembles at packing fractions above $54\%$. This notion of degeneracy differs from the degeneracy of a quasiperiodic random tiling arising through phason flips. Free energy calculations show that a triclinic crystal is preferred at high packing fractions.
\end{abstract}

\maketitle

Hard disks and spheres order into hexagonal and face-centered cubic crystals, respectively, above a certain packing fraction. A more complex phase behavior is observed if the disks or spheres are rigidly bonded into dimers (dumbbells)~\cite{WojciechowskiPRL1991, VegaJCP1992, MalanoskiJCP1997, MarechalPRE2008}. A solid phase, disordered in the orientation of dimers while ordered on the monomer level, forms if the distance between monomers within a dimer is roughly the diameter of a monomer. This equilibrium solid phase can be alternatively understood as a random pairing of neighboring monomers within the native monomer crystal. The resulting thermodynamic ensemble of ground states is degenerate and the structure is therefore called a degenerate crystal. As shown by Wojciechowski \emph{et al.}~\cite{WojciechowskiPRL1991} for hard disks, the entropy associated with the degeneracy exceeds the entropy from excluded volume effects, which by itself is sufficient to drive the crystallization of hard monomers. Other consequences of the pairing of monomers into dimers include topological defects~\cite{BluntScience2008}, a restricted, glassy dislocation motion~\cite{CohenPRL2008, GerbodeCohenPRL2010}, and unusual elastic properties~\cite{TretiakovJNCryst2006}. Similar degenerate phases have also been observed for freely-joined chains of hard spheres~\cite{KarayiannisLasoPRL2009, KarayiannisLasoSoftMatter2010}.

\begin{figure}
  \centering
  \includegraphics[width=\columnwidth]{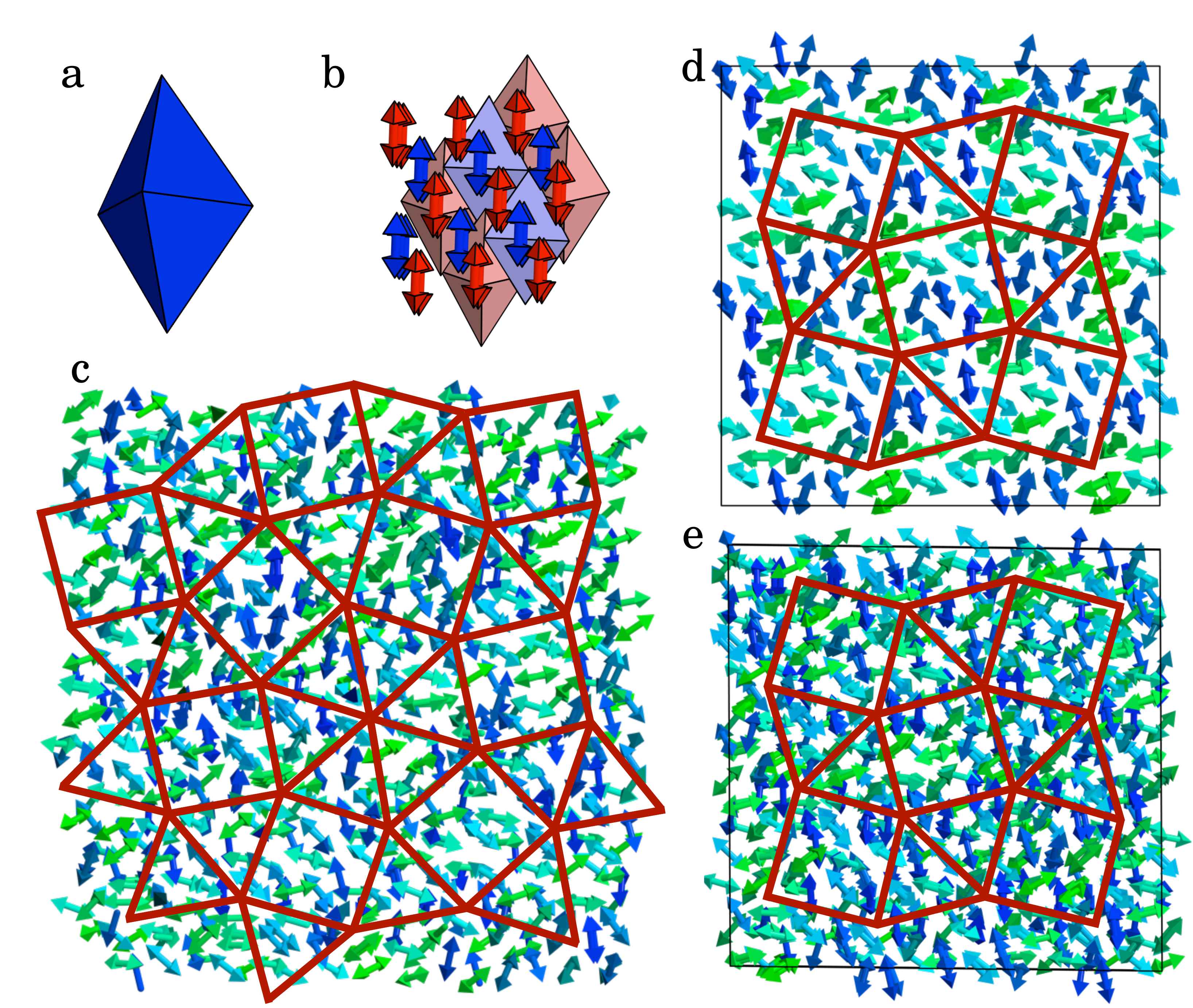}
  \caption{\label{fig:summary} Phases formed by (a) triangular bipyramids (TBPs): (b) TBP crystal, (c) degenerate quasicrystal, (d) regular quasicrystal approximant, (e) degenerate quasicrystal approximant. For visualization purposes, we show member tetrahedra of most TBPs at $30\%$ actual size and connect their centers with bonds. In (c-e), tetrahedra and bonds are colored according to their orientation projected on the plane.}
\end{figure}

Although degenerate crystals can potentially assemble from dimers of hard shapes other than disks and spheres, few examples have been reported. One reason is the competition between degenerate crystals and the liquid crystalline phases frequently observed for particles with large aspect ratios. For example, elongated tetragonal parallelepipeds, which for an aspect ratio of 2:1 can be viewed as dimers of face-sharing cubes, form a degenerate parquet phase at intermediate densities before transforming into a smectic liquid crystal that eventually crystallizes~\cite{JohnEscobedo2008}. Another simple dimer is the triangular bipyramid (TBP), which consists of two face-sharing, regular tetrahedra (Fig.~\ref{fig:summary}a). The TBP is the simplest face-transitive bipyramid and the twelfth of the $92$ Johnson solids. The lack of inversion symmetry of the TBP, however, makes lattice packings non-optimal~\cite{TorquatoJiaoNature2009}, and thus it is potentially more interesting as a dimer than dimers of spheres and cubes. Moreover, the recent synthesis of TBP-shaped nanoparticles and colloids~\cite{Wiley2006, JiangBullChemSocJpn2009, MirkinAngewChemIntEd2009, MirkinJACS2010} makes the investigation of this building block of practical relevance.

In both of the known ordered phases of hard, regular tetrahedra, each tetrahedron is in almost-perfect face-to-face contact with at least one other tetrahedron. The densest known packing of tetrahedra ($\phi=\frac{4000}{4671}\approx85.63\%$) is a parallel arrangement of two dimers (four tetrahedra) -- that is, two TBPs -- in a triclinic unit cell to form a dimer crystal~\cite{Chen2010,Kallus2010}, which we refer to in the present paper as the TBP crystal (Fig.~\ref{fig:summary}b). At lower packing fractions, hard tetrahedra assemble into a dodecagonal quasicrystal~\cite{HajiAkbariEtAl2009}, in which the tetrahedra form a decorated square-triangle tiling~\cite{OxborrowPRB1993}. 
Degenerate phases are impossible in the TBP crystal because the contacts between neighboring tetrahedra in different TBPs are highly imperfect~\cite{HajiAkbaricondmat2011}, but are possible in the quasicrystal due to the almost-perfect face-to-face contacts between all neighboring tetrahedra. Quasicrystals are solids with long-range order but without periodicity~\cite{LevinePRL1984}. Originally discovered in metallic alloys~\cite{Shechtman1984}, many alloy quasicrystals are now known, and a handful of quasicrystals have been reported in non-metallic systems. Among them are quasicrystals made from spherical micelles~\cite{Zeng2004}, binary nanoparticles~\cite{Talapin2009}, and hard tetrahedra~\cite{HajiAkbariEtAl2009}.

In this Letter, we investigate the phase behavior of hard TBPs and report a degenerate quasicrystal. The notion of degeneracy discussed here should not be confused with the extensively studied degeneracy associated with random tiling quasicrystals~\cite{ElserCommentPRL1985, StrandburgJaricPRL1989} where tiles with unique decoration patterns mix to form random tilings. We instead report a new type of randomness in the level of decorating individual tiles, in addition to the degeneracy of the random tiling. 

\begin{figure}
  \centering
  \includegraphics[width=\columnwidth]{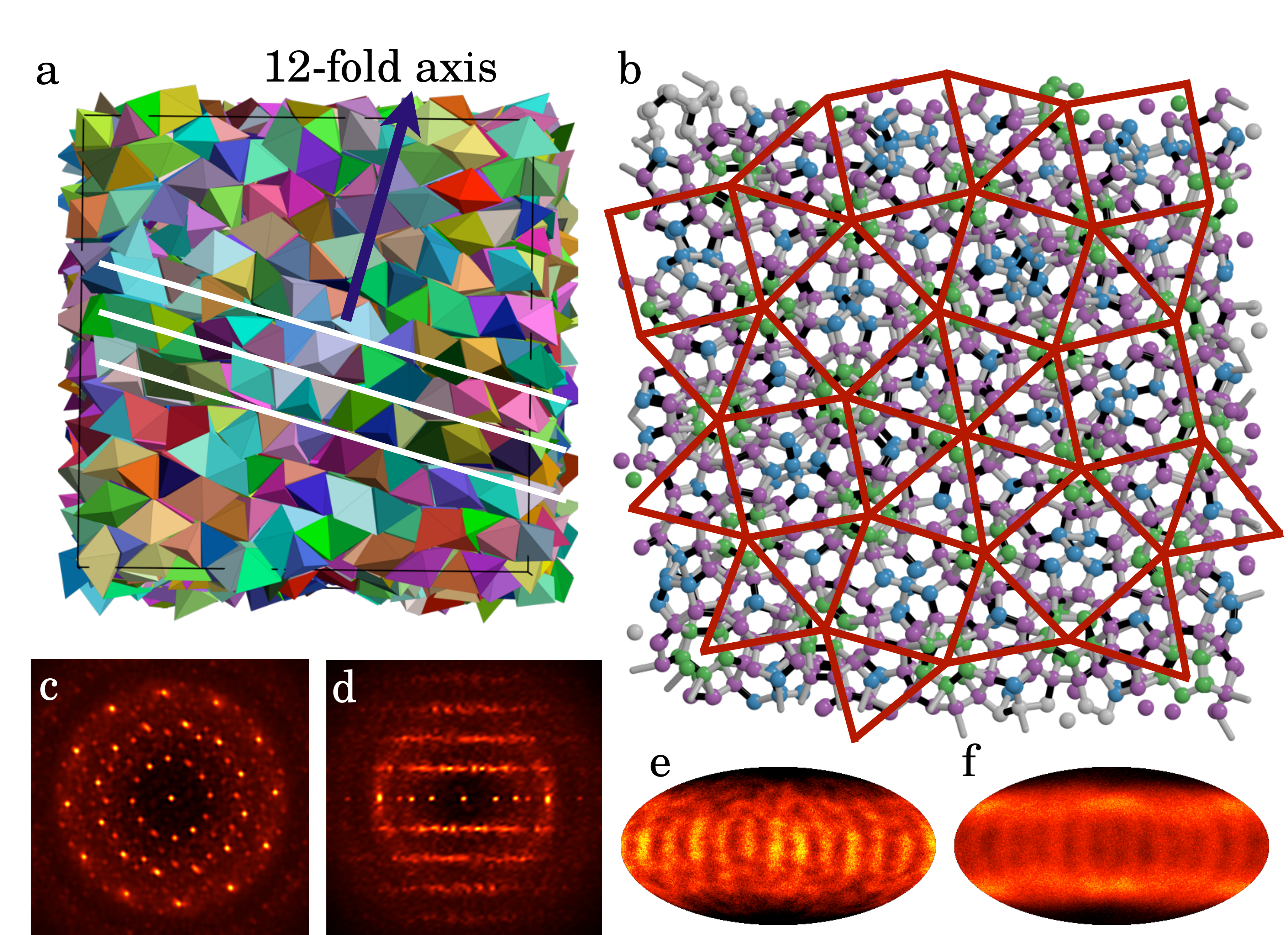}
  \caption{\label{fig:QC} (a) TBPs assemble into a dodecagonal quasicrystal in isobaric and isochoric Monte Carlo simulations. (b) The square-triangle tiling obtained by connecting the centers of 12-fold rings of member tetrahedra. Intra- and inter-TBP bonds are depicted in black and gray respectively. (d,e) Diffraction patterns with centers of member tetrahedra as scatterers calculated (c) perpendicular to and (d) across the layers. (e) Intra-TBP and (f) total bond order diagrams.}
\end{figure}

%%%%% METHODS

We use isochoric and isobaric Monte Carlo (MC) simulations to study hard TBPs, which we model as perfect polyhedra with sharp vertices and edges of unit length $\sigma$. Simulations are carried out within periodic boxes with system sizes ranging from $432$ to $8,\!000$ particles. Each isochoric MC cycle comprises one update per particle on average, which is either a trial translation or a trial rotation with equal probabilities. An additional box trial move is included per isobaric cycle. For fluid phases, the box is resized isotropically only, while for crystals its shape is also allowed to fluctuate. Free energies are calculated using thermodynamic integration and a modified Frenkel-Ladd method~\cite{FrenkelLaddJCP1984, VegaJCP1992} as described in detail in \cite{HajiAkbaricondmat2011}. Further details and simulation parameters are given in Ref.~\footnote{See supplementary material for more details.}.

%%%%% QUASICRYSTAL

The dodecagonal quasicrystal of TBPs forms spontaneously from the equilibrium fluid phase at packing fractions above $54\%$. Fig.~\ref{fig:QC}a depicts a side view of the quasicrystal formed in an isobaric simulation of $2,\!624$ TBPs at reduced pressure $P^*\!=\!P\sigma^3/k_BT\!=\!46$ and subsequently compressed to a packing fraction of $81.34\%$. TBPs arrange into layers (white lines), which stack on top of each other perpendicular to the 12-fold symmetry axis (dark arrow). We confirmed that the formation of the quasicrystal occurs reproducibly in systems with at least a few thousand particles and does not depend on the shape of the simulation box.

The quasicrystal structure can be best understood by replacing each bipyramid by its two member tetrahedra. Fig.~\ref{fig:QC}b depicts the centroids of tetrahedra within a few layers of Fig.~\ref{fig:QC}a. Neighboring tetrahedra are connected with bonds~\footnote{Two tetrahedra are defined as neighbors if their distance lies within the first peak of $g_t(r)$, the radial distribution function based on the centroids of the tetrahedra.}. Dodecagons that are depicted in purple in Fig.~\ref{fig:QC}b correspond to rings of twelve member tetrahedra, a structural motif characteristic of the quasicrystal~\cite{HajiAkbariEtAl2009}. These rings are further capped with pentagonal dipyramids (PDs), five tetrahedra sharing an edge, visible in the figure as pentagons (green) within dodecagons. Additional member tetrahedra, referred to as interstitials, fill the space between the rings and are depicted in dark blue. Together, dodecagons and PDs form motifs whose centers are the vertices of square and triangle tiles. Their mixing gives the square-triangle tiling its overall twelve-fold symmetry as observed in the diffraction pattern depicted in Fig.~\ref{fig:QC}c. Layering along the 12-fold axis can be seen in Fig.~\ref{fig:QC}d. Overall, the arrangement of the member tetrahedra is identical to that reported in the hard tetrahedron system~\cite{HajiAkbariEtAl2009}.

To elucidate how the bipyramids are arranged within the quasicrystal, we compare statistical distributions of intra-TBP bonds (bonds that connect member tetrahedra within TBPs) and the set of all bonds in the quasicrystal by projecting both sets onto the surface of a unit sphere. The resulting diagrams are referred to as intra-TBP and total bond order diagrams, respectively and are visualized using the Mollweide projection with the $12$-fold axis pointing in the vertical direction. Comparing these bond order diagrams (Figs.~\ref{fig:QC}e,f), we observe no significant difference in the distribution of bond directions within the 12-fold layers. This suggests that the pairing of tetrahedra in the quasicrystal does not follow a predefined set of rules and is instead random. However, tetrahedra tend to pair more strongly within layers than between neighboring layers, a fact that can be explained by noting that face-to-face contacts are more perfect within layers. Motivated by studies of hard sphere dimers~\cite{WojciechowskiPRL1991}, we refer to the TBP quasicrystal as a degenerate quasicrystal (DQC). The randomness can be seen clearly in Fig.~\ref{fig:summary}c. It is surprising that the structural quality of the DQC is uncompromised despite the additional geometrical constraints imposed on the system by pairing tetrahedra into TBPs. For instance, we find that the maximum packing fraction achieved by replacing the bipyramids with individual member tetrahedra and then compressing is statistically identical to that obtained in simulations of hard tetrahedra.

%%%%% APPROXIMANTS

\begin{figure}
  \centering
  \includegraphics[width=\columnwidth]{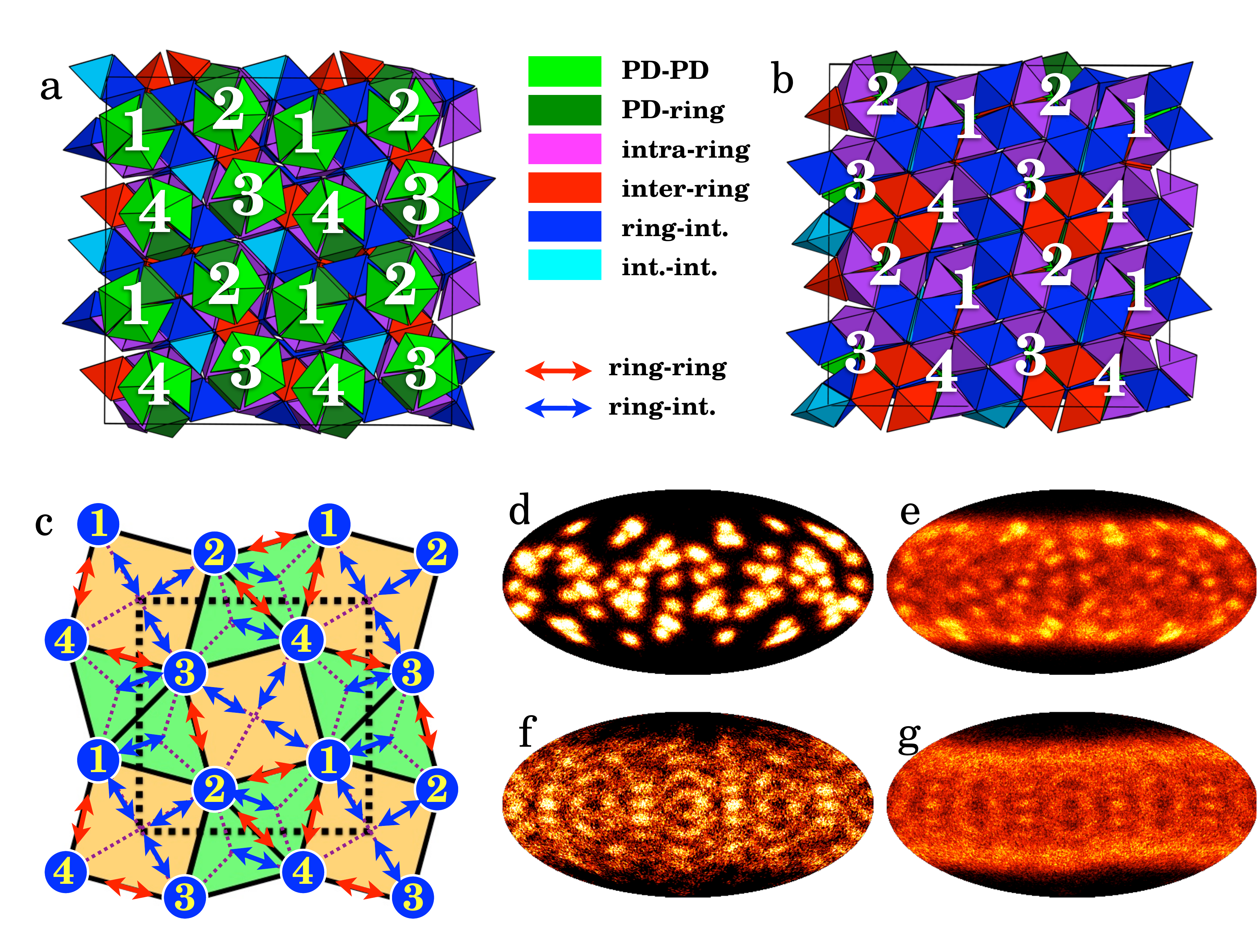}
  \caption{\label{fig:approximant}(a)~Top and (b)~bottom views of the regular approximant. The unit cell has $41$ triangular bipyramids. Particles are colored according to their environment: PD-PD (light green), PD-ring (dark green), intra-ring (purple), ring-ring (red), ring-interstitial (blue), interstitial-interstitial (cyan). (c)~Schematics of the unit cell with connections between neighboring rings and between rings and central interstitials shown with red and blue double-arrows, respectively. (d,f)~Intra-TBP and (e,g)~total bond order diagrams for (d,e)~the regular approximant and (f,g)~the degenerate approximant. In the legends, `int' stands for `interstitial'.}
\end{figure}

Approximants are periodic phases that are structurally similar to the quasicrystal locally~\cite{GoldmanKeltonRevModPhys1993}. Constructing an approximant of the TBP quasicrystal involves not only choosing a periodic tiling and decorating it with tetrahedra, but also pairing the tetrahedra into bipyramids. We choose the $(3.4.3^2.4)$ Archimedean tiling which, in the case of hard tetrahedra, gives rise to the densest approximant~\cite{HajiAkbariEtAl2009}. There is no unique way of pairing tetrahedra into TBPs even within a single unit cell of the approximant due to degeneracies associated with rotations of the capping PDs. In particular, it is not possible to avoid breaking the four-fold symmetry of the approximant unit cell in the pairing process. We constructed a \emph{regular approximant} by retaining as much of the symmetry as possible. Top and bottom views of the constructed approximant are depicted in Figs.~\ref{fig:approximant}a,b while a unit cell is depicted in Fig.~\ref{fig:approximant}c where ring-ring and ring-interstitial connections are highlighted. We find that the regular approximant can be compressed to a maximum packing fraction of $83.39\%$, a bit less than the maximum packing fraction of $85.03\%$ achieved for the quasicrystal approximant constructed of individual tetrahedra~\cite{HajiAkbariEtAl2009}. The distinctive difference between the intra-TBP (Fig.~\ref{fig:approximant}d) and the total bond order diagrams (Fig.~\ref{fig:approximant}e) is a result of this deterministic pairing (Fig.~\ref{fig:summary}d). 

By expanding the regular approximant, we find that it melts at $P^*\le35$ and packing fractions $\phi <54\%$. But before melting, the crystal slowly transforms into a more loosely packed structure in which tetrahedra are paired at random into TBPs, just as in the DQC, although their positions and orientations are unchanged (Fig.~\ref{fig:summary}e). The resulting structure is therefore degenerate to the tetrahedron-based approximant and we refer to it as a degenerate approximant (DA). The angular distribution of intra-TBP bonds around the four-fold axis (Fig.~\ref{fig:approximant}f] is more similar to that of all bonds (Fig.~\ref{fig:approximant}g] in the degenerate approximant than in the case of the regular approximant [Figs.~\ref{fig:approximant}d,e], which again suggests random pairing. We find that the transformation from regular to degenerate approximant is irreversible on the time scale of our simulations ($\approx10^8$ MC cycles). Since the DA can only be recompressed to a density of $82.88\%$, which is lower than the maximum density of the regular approximant, the DA has to be stabilized by its pairing disorder close to melting.

\begin{figure}
  \centering \includegraphics[width=\columnwidth]{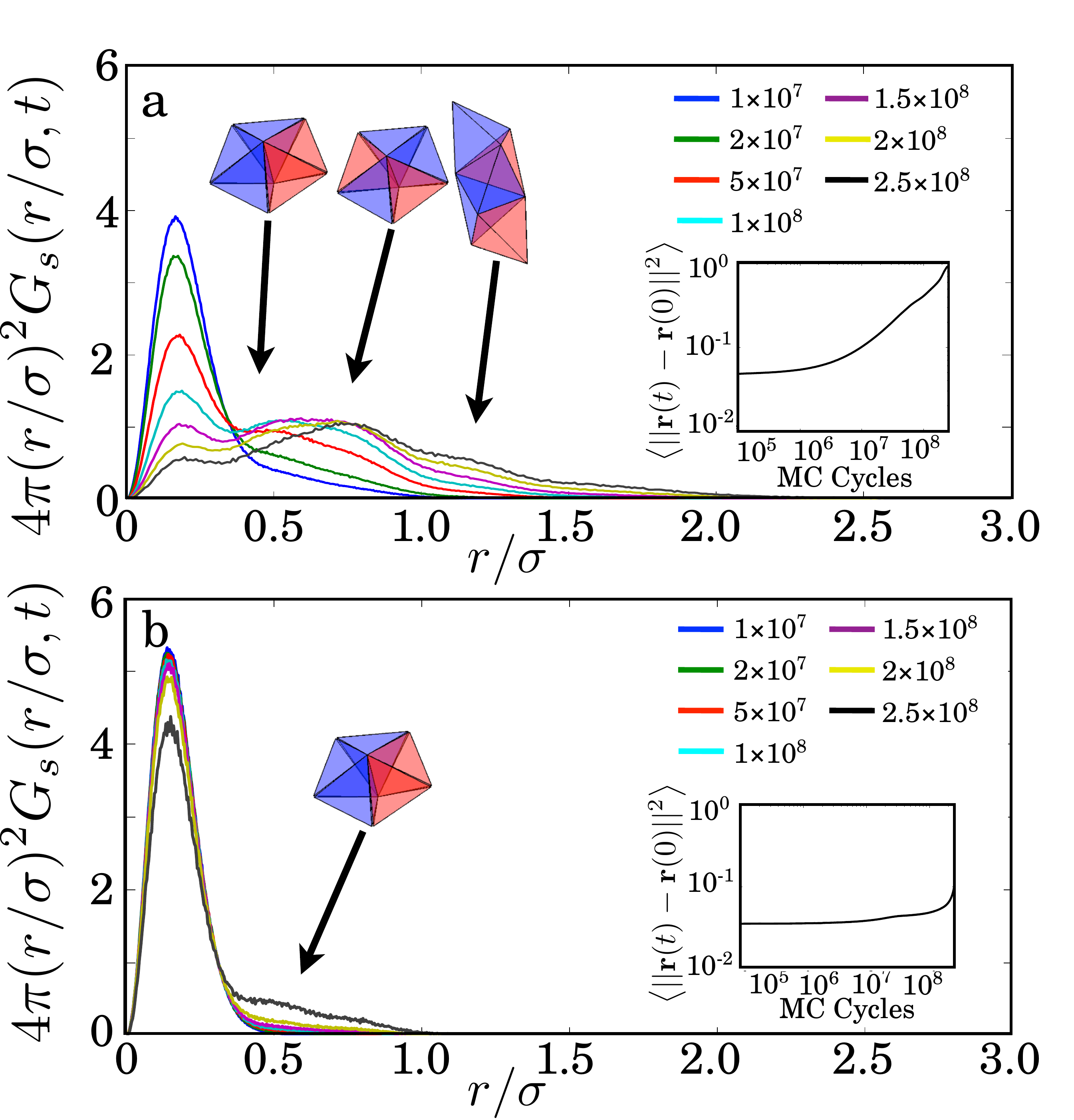}
  \caption{\label{fig:app_dynamics}The self part of the van Hove correlation function $G_s(r,t)$ measures the particle motion in the approximant. The separation distance $r(t)$ is calculated between centers of mass of member tetrahedra. (a) Large rearrangements occur at $\phi=57\%$. (b) There is little motion present at $\phi=60\%$. The observed dynamics is similar to that observed in the hard tetrahedron system~\cite{HajiAkbaricondmat2011}.}
\end{figure}

To understand how the regular approximant transforms into the DA, we note (Fig.~\ref{fig:QC}b) that the arrangement of the member tetrahedra can be alternatively understood as a spanning network of interpenetrating PDs~\cite{HajiAkbariEtAl2009}. In the hard tetrahedron system, PDs can easily rotate around their principal axes~\cite{HajiAkbaricondmat2011}. Such rotations are also essential in understanding the local rearrangements of bipyramids at densities below $60\%$.  As shown in Fig.~\ref{fig:app_dynamics}b, TBPs move very little at $\phi=60\%$. Even after $250$ million MC cycles only a small fraction of TBPs have moved as much as $\sigma$. A much faster dynamics occurs at $\phi=57\%$. Particles at or near that density move over discrete distances that are characteristic of a PD network (Fig.~\ref{fig:app_dynamics}a). These rearrangements change neither the tiling nor its decoration. Instead, they reshuffle the pairing pattern by a sequence of PD rotations. After a sufficiently large number of reshuffling moves the DA emerges from the regular approximant.

%%%%% FREE ENERGY CALCULATIONS

\begin{figure}
  \centering
  \includegraphics[width=\columnwidth]{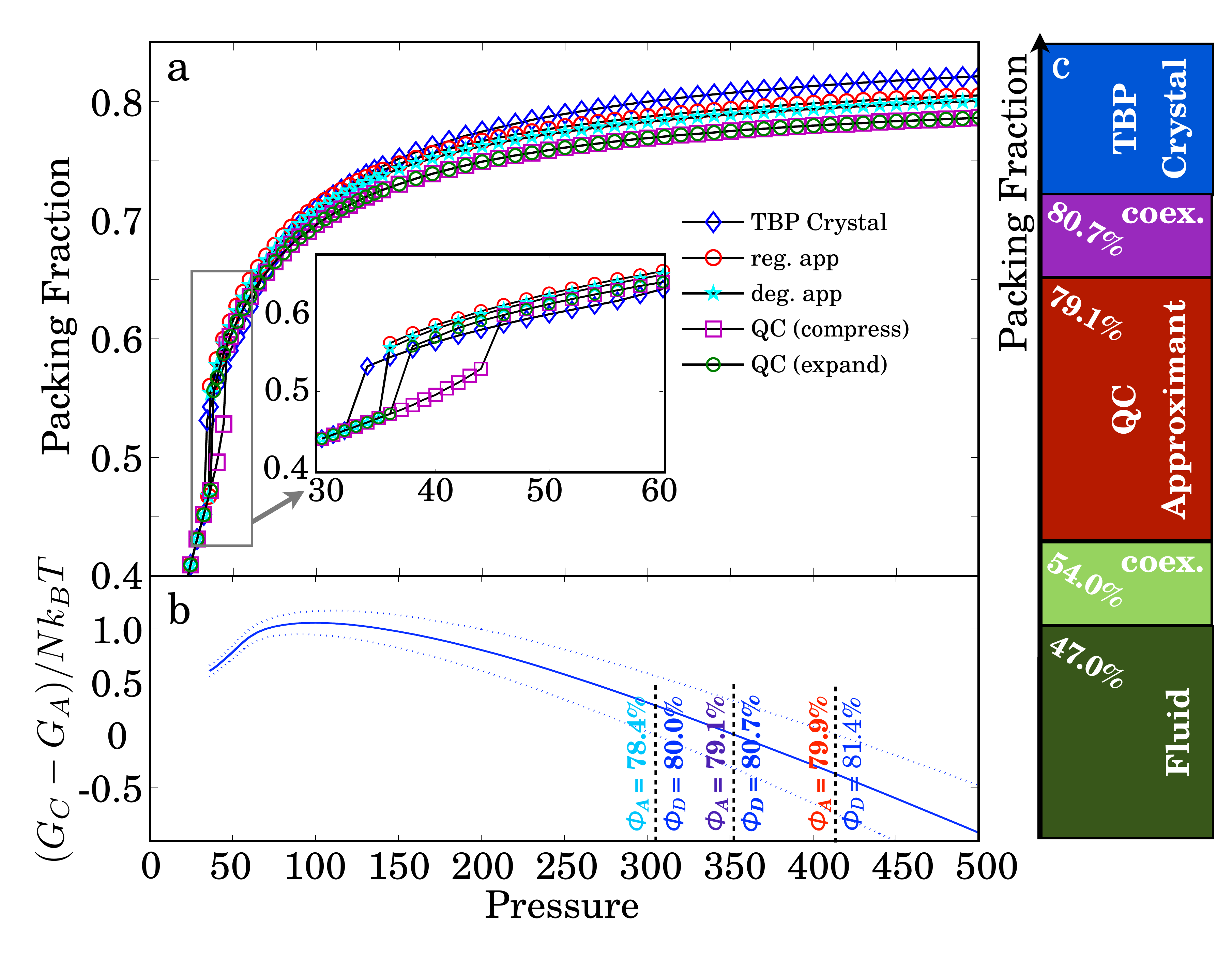}
  \caption{\label{fig:thermodynamics}(a) Equation of state for the TBP crystal, the degenerate quasicrystal, the regular and the degenerate approximants. (b) The free energy difference between the TBP crystal and the approximant. (c) Equilibrium phases of hard TBPs.}
\end{figure}

Next we study the relative thermodynamic stability of various phases. We first compare the DQC obtained in simulation and its constructed approximants. As observed in Fig.~\ref{fig:thermodynamics}a, both the regular and the degenerate approximant are slightly denser than the DQC at all pressures. The relation $G(P^*_2)-G(P^*_1)\propto\int_{P^*_1}^{P^*_2}\phi^{-1}\text{d}P^*$ between the free energy and the equation of state then suggests that the approximants are thermodynamically preferred over the DQC at sufficiently high pressures because their Gibbs free energies increase more slowly with pressure. Furthermore, the approximants melt at lower pressures than the quasicrystal, which indicates that they might even be more stable than the quasicrystal at all pressures. Nevertheless, the DQC remains the only ordered phase that forms in our simulations. It is also the only structure we expect to be observed in experiments of hard nanocolloidal TBPs since the kinetic process of transforming from the DQC into the approximant is extremely slow. Considering the local structural similarity of the DQC and the fluid in terms of the PD network, the formation of the less stable DQC and not the approximant in simulation may be another example of Ostwald's rule~\cite{FrenkelOstwaldRule1999}.

Next, we compare the approximant with the TBP crystal by calculating the free energy difference between them. As shown in Fig.~\ref{fig:thermodynamics}b, the approximant has a lower free energy than the TBP crystal for packing fractions below $79\%$. A phase transition occurs at $P^*_c=356\pm50$, corresponding to coexistence packing fractions of $\phi_{c,\text{app}}=(79.1\pm0.8)\%$ and $\phi_{c,\text{TBP}}=(80.7\pm0.7)\%$. The thermodynamic stability of the approximant at lower densities can be attributed to the additional configurational entropy associated with collective motions of particles. Such motions are not present in the TBP crystal. Their role in stabilizing the quasicrystal approximant has been shown for the structurally and dynamically similar system of hard tetrahedra~\cite{HajiAkbaricondmat2011}. The phase diagram of the hard TBP system is depicted in Fig. \ref{fig:thermodynamics}c.

Remarkably, hard TBPs not only prefer a complex quasicrystal over the simpler TBP crystal at intermediate packing fractions, but also form it on timescales comparable to that previously observed in the hard tetrahedron system. This is surprising because, in comparison to tetrahedra, the motion of the highly anisotropic bipyramids is considerably more constrained. Nevertheless, the degeneracy of the quasicrystal helps it form easily in simulation. Random pairing allows TBPs to join existing seeds of the DQC without forming configurations that are kinetically trapped due to incorrect pairing. Particle rearrangements needed for the formation and growth of the seed are also feasible due to the local similarity of the fluid and the quasicrystal~\cite{KeysGlotzer2007, HajiAkbariEtAl2009}. Finally, the degeneracy and the existence of ring-ring and ring-interstitial "cross-links" adds rigidity to the TBP structures. This means that the TBP system might be superior over the tetrahedron system in terms of its mechanical properties, just as for crystals of hard sphere dimers compared to crystals of their monomers~\cite{TretiakovJNCryst2006}. 

%%%%% CONCLUSIONS

In conclusion we have shown that hard triangular bipyramids form a degenerate dodecagonal quasicrystal. Our finding is only the second quasicrystal formed with hard particles, the first reported degenerate quasicrystal, and one of only a few quasicrystals formed in non-atomistic systems. Our results suggest that degenerate phases are not restricted to simple close-packed crystals and might be common in dimer systems.

%%%%% ACKNOWLEDGMENTS

This work was supported in part by the U. S. Air Force Office of Scientific Research (FA9550-06-1-0337) and by a U.S. Department of Defense National Security Science and Engineering Faculty Fellowship (N00244-09-1-0062). M.E. acknowledges support from the Deutsche Forschungsgemeinschaft.  A.H.-A. acknowledges support from the University of Michigan Rackham Predoctoral Fellowship program.

\bibliographystyle{apsrev}
\bibliography{HardTetBib}

%\eject\phantom{}
\newpage

\appendix 
\section{SUPPLEMENTARY INFORMATION}

\setcounter{figure}{0}
\renewcommand{\thefigure}{S\arabic{figure}}
\renewcommand{\thetable}{S\arabic{table}}

\section{Compression Algorithm}
To overcome the sluggishness of conventional isobaric Monte Carlo (MC) simulations in compressing hard particle systems with large numbers of particles, we utilize a modified compression algorithm described in detail in~\cite{HajiAkbariEtAl2009} for obtaining the densest packings of the quasicrystal and the approximants. This method relies on allowing a small number of minor overlaps in compression moves, which are subsequently removed to obtain dense packings. All expansion/compression moves are accepted even if they result in new overlaps and a different criterion is used to ensure that the number and amount of overlaps remains small. We monitor $p$, the fraction of trial translations accepted since the previous volume move, and compare it to a target acceptance probability $p_t$. If $p<p_t$, the volume of the box is increased by a random factor $f$ uniformly chosen from the interval $[1,1+0.002\Delta{x}]$ while for $p>p_t$ a compression is attempted by a random factor uniformly distributed in $[1-0.002\Delta{x},1]$. Between volume moves, the system evolves through conventional trial translations and trial rotations and the moves that do not generate new overlaps are accepted. $\Delta{x}$, the maximum step size for a trial translation is the control parameter in this method and is inversely related to pressure in the conventional isobaric MC algorithm. In a compression run, $\Delta{x}$ is exponentially decreased until the densest packing is obtained. A few hundred cycles of conventional NVT MC are carried out at the end to remove any overlaps generated by this method. 

\section{Free energy calculations}

We use a modified version of the Frenkel-Ladd thermodynamic integration scheme~\cite{FrenkelLaddJCP1984, VegaJCP1992} to calculate the Helmholtz free energy differences between various crystals in the system. The Gibbs free energy is then determined from the Helmholtz free energy together with the equation of state.  The equation of state is also used to extrapolate the Gibbs free energies to pressures where no Frenkel-Ladd calculation can be performed by integrating $\text{d}G=V\text{d}P$. The values of $\gamma_{\max}$ and $c$ chosen in this study are included in Table~\ref{table:system_sizes}. Further technical details of the free energy calculation scheme can be found in~\cite{HajiAkbaricondmat2011}, where the same method was applied to a system of hard tetrahedra.

\section{Van Hove correlation function and structure factor}

The van Hove correlation function and the structure factor are calculated for centroids of member tetrahedra in the hard TBP quasicrystal approximant. To determine the structure factor, the centroids are convoluted with a Gaussian and projected along the observation direction. The resulting pattern is sheared into a square, discretized, fast Fourier transformed, and then sheared back.

\section{Correlated Motions and Additional Entropy}
The entropy of a hard particle system is the volume of the $6N$-dimensional configuration space accessible to it. In dense arrangements where particles rattle in their respective cages, this accessible volume can be approximated by the product of free volumes of individual particles with the free volume being the volume of the $6$-dimensional configurational space of a given particle accessible to it while all other particles are kept fixed (mean-field approximation~\cite{VegaMonsonMolPhys1992, HajiAkbaricondmat2011}). Through collective motions, particles can access regions of the phase space not accessible through rattling of individual particles. Therefore collective motions contribute to some additional entropy not accounted for by the mean-field approximation and further stabilize the corresponding system. The stabilizing effect of correlated motions in the quasicrystal approximate was rigorously confirmed for the hard tetrahedron system by comparing exact free energies calculated from the Frenkel-Ladd method with the free energies estimated from calculated free volumes and the mean-field approximation~\cite{HajiAkbaricondmat2011}. A comparison of Fig. 4 in the present work with Fig. 9 in~\cite{HajiAkbaricondmat2011} demonstrates the remarkable similarity of the dynamics in the TBP and tetrahedron quasicrystal approximants, in particular with regards to collective motions of the type described above. We therefore expect these collective motions to have similar stabilizing effects in both systems.

\section{Dense packings}

The densest unit cells of the TBP crystal and the regular approximant are given in Tables~\ref{table:dense_TBP} and~\ref{table:dense_reg_app}, respectively. In the tables, the orientation of each TBP is represented by a unit quaternion $\textbf{q}=(q_t,q_x,q_y,q_z)$ with rotation matrix
\begin{eqnarray*}
\left(
	\begin{array}{ccc}
		q_t^2+q_x^2-q_y^2-q_z^2 & 2(q_xq_y-q_tq_z) & 2(q_xq_z+q_tq_y) \\
		2(q_xq_y+q_tq_z) & q_t^2-q_x^2+q_y^2-q_z^2 & 2(q_yq_z-q_tq_x) \\
		2(q_xq_z-q_tq_y) & 2(q_yq_z+q_tq_x) & q_t^2-q_x^2-q_y^2+q_z^2
	\end{array}
\right)
\end{eqnarray*}
The vertices of the TBP with $\textbf{q}=(1,0,0,0)$ are given by:
\begin{eqnarray*}
\textbf{v}_{1,2} &=& \left(0,0,\pm{4\sqrt3}/{3}\right),
\textbf{v}_{3,4} = \left(-\sqrt6/{3},\pm\sqrt2,0\right),\\
\textbf{v}_5 &=& \left({2\sqrt6}/{3},0,0\right).
\end{eqnarray*}

%\newpage

\begin{table*}
	\caption{\label{table:system_sizes}Simulation details for the calculation of the equation of state (EOS), quasicrystal (QC) assembly, and thermodynamic integration. Several independent runs were performed for each state point to assure accurate statistics. The smallest system used for quasicrystal formation has $1,\!458$ particles. Quasicrystal formation is robust and is routinely observed at densities above $54\%$ and system sizes larger than $1,\!400$ particles. We do not observe a quantitative difference in the structure of the quasicrystal for such large system sizes. The small, $432$-particle system is only used for the mathematically constructed TBP crystal (with 216 two-particle unit cells), which is used for estimating the equation of state and free energy calculations only.}
	\begin{tabular}{llcccc}
		\hline\hline
		~~~Phase~~~ & ~~~Objective~~~ &Ensemble & System size & MC sweeps & Parameters\\
		\hline
		Fluid & EOS calculation & isobaric & $2,\!624$ & $10^7$ & $0.01\le P^*\le60$ \\
		Fluid & QC assembly & isobaric & $1,\!458-2,\!624$ & $10^8$ & $40\le P^*\le60$\\
		Fluid & QC assembly & isochoric & $8,\!000$ & $10^8$ & $0.5\le\phi\le0.6$\\
		TBP crystal & EOS calculation & isobaric & $432$ & $10^7$ & $32\le P^*\le 10,000$\\
		Quasicrystal & EOS calculation & isobaric & $2,\!624$ & $10^7$ & $36\le P^*\le 10,000$ \\
		Approximant & EOS calculation & isobaric & $656$ & $10^8$ & $35\le P^*\le 10,000$\\ 
		Approximant & Dynamics & isochoric & $656$ & $3\times10^8$ & $0.57\le\phi\le0.80$\\
		TBP crystal & Therm. Integration & isochoric & $432$ & $2\times10^5$ per $\gamma$ & $\begin{array}{c}0\le\gamma\le4\times10^6,c=1/2\\ 0.70\le\phi\le0.80\end{array}$ \\
		Approximant & Therm. Integration & isochoric & $656$ & $2\times10^5$ per $\gamma$ & $\begin{array}{c}0\le\gamma\le4\times10^6,c=1/2\\ 0.60\le\phi\le0.80 \end{array}$ \\
		\hline\hline
	\end{tabular}
\end{table*}

\begin{table*}
	\caption{\label{table:dense_TBP}A unit cell of the TBP crystal with $\phi=85.6347\%$. The lattice vectors are given by $b_1=(8/5, 12/5, 4/5), b_2=(3/8,97/80, 191/80), b_3=(12/5, 3/20, 3/2)$. The positions and quaternions of the two particles in the unit cell are given.}
	\begin{tabular}{l|ccc|rrrr}
	\hline
	\hline
	$i$ & $x_i$ & $y_i$ & $z_i$ & $q_{t,i}$ & $q_{x,i}$ & $q_{y,i}$ & $q_{z,i}$ \\
	\hline
	$1$& $0$	&$0$ &	$0	$ &$0.880476$ &	$-0.364705$	& $0.279848$ &	$-0.115917$ \\
$2$ &	$-26/15$ &	$-79/120$ &	$-1/120$	& $-0.279848$ &	$0.115917$  & $0.880476$ &	$-0.364705$ \\
	\hline\hline
	\end{tabular}
\end{table*}

\begin{table*}
	\caption{\label{table:dense_reg_app}A unit cell of the regular approximant with $\phi=83.39\%$. Lattice vectors are given by $b_1=(9.96864731,0,0), b_2=(0,10.12720286,0), b_3=(0,0,2.597424124)$. There are $41$ particles in the unit cell; their positions and orientations are given.}
	\begin{tabular}{l|rrr|rrrr}
	\hline
	\hline
	$i$ & $x_i$ & $y_i$ & $z_i$ & $q_{t,i}$ & $q_{x,i}$ & $q_{y,i}$ & $q_{z,i}$ \\
	\hline
1	&$-3.086054086$&	$-2.363988414$&	$0.413277757$&	$0.724311194$&	$0.366868384$&	$0.073525829$& $0.579115564$ \\ 
2	&$-4.118404870$&	$-0.892236112$&	$0.415845679$&	$0.432135294$&	$-0.510211566$&	$0.631248159$& $0.393025455$ \\ 
3	&$-3.924682848$&	$-3.466017697$&	$-0.350811567$&	$-0.318217496$&	$0.434494644$&	$0.725846175$& $0.427901111$ \\ 
4	&$-0.462122468$&	$-4.240376441$&	$-0.782025752$&	$0.852598155$&	$0.387765410$&	$-0.303556385$&	$0.174836762$ \\ 
5	&$-0.327948406$&	$-1.568864202$&	$-1.104708664$&	$0.131985547$&	$-0.297203708$&	$-0.551333205$&	$0.768297773$ \\ 
6	&$-2.672857633$&	$0.001417971$&	$-0.337285874$&	$-0.555926418$&	$-0.431797813$&	$0.706493586$& $-0.073234406$ \\ 
7	&$-2.069478657$&	$-1.471298448$&	$-0.432573444$&	$0.177650291$&	$0.552837588$&	$0.096194039$& $0.808429145$ \\ 
8	&$-0.826248624$&	$-0.277415522$&	$-0.173475293$&	$0.812731912$&	$0.302601899$&	$-0.360382520$&	$-0.343545295$ \\ 
9	&$-2.318916062$&	$-4.112336247$&	$0.659756236$&	$-0.410542691$&	$-0.392643954$&	$0.520886557$& $0.637151959$ \\ 
10	&$-1.344851368$&	$-2.859967477$&	$0.633186749$&	$0.636896655$&	$0.725358601$&	$0.219148044$& $0.142097453$ \\ 
11	&$-4.696641186$&	$1.276225643$&	$0.602240973$&	$0.360108299$&	$0.252500774$&	$0.269483792$& $0.856705234$ \\ 
12	&$-3.747047319$&	$5.055281040$&	$-1.103638744$&	$0.780999587$&	$-0.449388327$&	$0.430045356$& $-0.056131712$ \\ 
13	&$-0.058997540$&	$4.025246097$&	$-0.779658046$&	$-0.218095268$&	$0.293099666$&	$0.406142214$& $0.837601063$ \\ 
14	&$-0.019924758$&	$1.968363142$&	$-0.681742157$&	$-0.575391179$&	$0.092652143$&	$0.548061557$& $0.599974250$ \\ 
15	&$-2.649483230$&	$4.205046998$&	$-0.356921625$&	$0.507783250$&	$-0.424324513$&	$0.582993354$& $0.471406013$ \\ 
16	&$-1.457204318$&	$4.260893847$&	$1.190217106$&	$0.729944614$&	$-0.577104503$&	$0.065356154$& $0.360360689$ \\ 
17	&$-1.778009861$&	$2.641765034$&	$-0.715619163$&	$0.666191766$&	$0.458836083$&	$-0.016867546$&	$0.587684836$ \\ 
18	&$-4.267156145$ &	$3.386854105$ &	$-0.471378412$ &	$-0.000224002$ &	$0.782668146$&	$0.478172787$& $0.398473723$ \\ 
19	&$-1.197762201$&	$1.002543156$&	$-1.100736732$&	$0.473332539$&	$-0.577185196$&	$-0.123466505$&	$0.653888048$ \\ 
20	&$-2.944126080$&	$1.001577617$&	$0.918738785$&	$0.691606294$&	$0.561167807$&	$-0.315127681$&	$-0.327820027$ \\ 
21	&$-3.431835202$&	$2.531003145$&	$0.767326240$&	$0.489058954$&	$0.502023864$&	$0.569884742$& $0.428981072$ \\ 
22	&$4.166061318$&	$-1.755893743$&	$0.614482584$&	$0.857339372$&	$0.268730116$&	$-0.248620481$&	$-0.361857958$ \\ 
23	&$4.707576086$&	$-0.238559740$&	$-0.536923289$&	$0.079504631$&	$-0.419075935$&	$-0.566678392$&	$0.704932603$ \\ 
24	&$4.456114027$&	$-2.994208210$&	$-0.342843770$&	$0.780466896$&	$0.401090725$&	$-0.425640796$&	$0.220969608$ \\ 
25	&$4.827932349$&	$-4.702316768$&	$-0.159021811$&	$-0.126511507$&	$0.202433383$&	$0.414721860$&	$0.878078210$ \\ 
26	&$3.567745446$&	$-4.594666250$&	$-1.082778182$&	$0.754021412$&	$-0.607585046$&	$-0.001432754$&	$0.249579784$ \\ 
27	&$1.938503081$&	$-3.805733080$&	$-0.665797086$&	$-0.348135133$&	$-0.057177952$&	$0.878347065$&	$0.322550839$ \\ 
28	&$0.916420109$&	$-4.715357725$&	$1.157640853$&	$-0.342223016$&	$-0.075509837$&	$-0.566097787$&	$0.746133344$ \\ 
29	&$0.595676354$&	$-2.632461476$&	$-0.391055241$&	$0.686531636$&	$-0.073835559$&	$0.706109295$&	$0.156946761$ \\ 
30	&$1.361780627$&	$-1.041508444$&	$-0.497237432$&	$-0.218226841$&	$0.281976797$&	$0.279418344$&	$0.891510808$ \\ 
31	&$2.522729926$&	$-1.366975480$&	$0.913207732$&	$0.340971281$&	$0.338907933$&	$0.555567249$&	$0.678398872$ \\ 
32	&$2.666309940$&	$-2.950705870$&	$0.792567704$&	$0.396438787$&	$0.527197369$&	$-0.529906306$&	$-0.533008939$ \\ 
33	&$2.576140986$&	$2.039647502$&	$0.467914212$&	$-0.571761301$&	$-0.070597781$&	$0.373290369$&	$0.727158351$ \\ 
34	&$3.524756753$&	$0.477954765$&	$0.412277663$&	$0.350644500$&	$0.670036628$&	$0.463160891$&	$-0.462148614$ \\ 
35	&$3.992316141$&	$3.709275417$&	$0.053283182$&	$0.183126889$&	$0.789653300$&	$0.085915702$&	$-0.579250119$ \\ 
36	&$4.199227190$&	$2.276875385$&	$-0.600510380$&	$-0.359557729$&	$0.681364585$&	$0.265991831$&	$-0.579403908$ \\ 
37	&$2.385703449$&	$4.617199613$&	$-0.648455546$&	$0.236634396$&	$-0.425417036$&	$0.711443254$&	$-0.506826404$ \\ 
38	&$0.932212413$&	$0.559919170$&	$0.041230498$&	$0.541559035$&	$-0.498432529$&	$0.611824825$&	$0.289739900$ \\ 
39	&$1.870351885$&	$3.740201639$&	$0.686805299$&	$0.635673515$&	$0.523964733$&	$0.374951541$&	$0.425195818$ \\ 
40	&$0.852464474$&	$2.539550949$&	$0.649899985$&	$-0.138338952$&	$-0.219309086$&	$0.717949504$&	$0.645998737$ \\ 
41	&$2.376730451$&	$0.544997559$&	$-0.642579527$&	$-0.670448515$&	$0.706351286$&	$-0.038435571$&	$-0.223806515$ \\ 
 \hline\hline
	\end{tabular}
\end{table*}

\begin{figure*}
	\includegraphics[width=2\columnwidth]{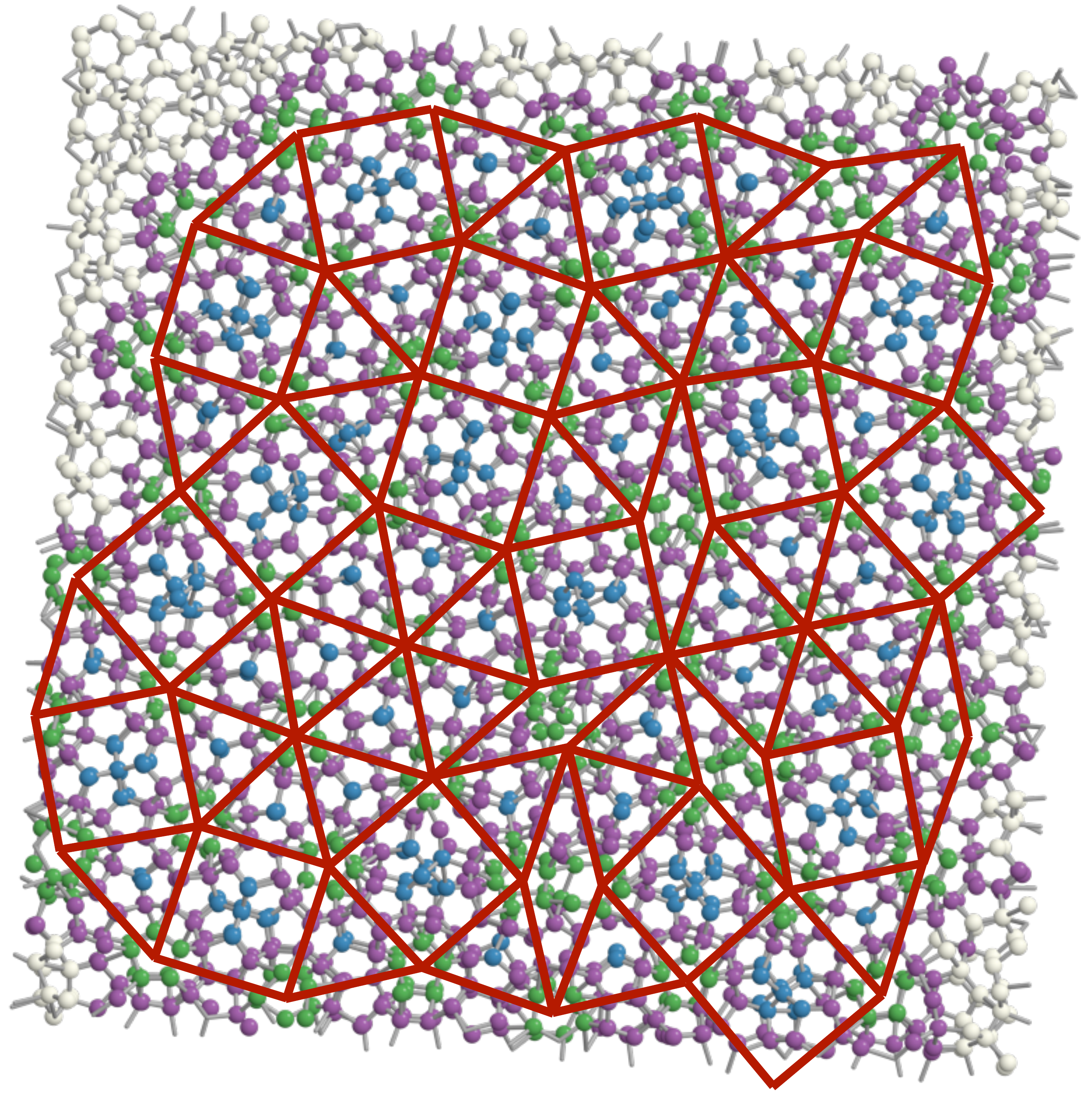}
	\caption{Tiling obtained for a slab of the degenerate quasicrystal formed in an NVT simulation of $8,\!000$ particles at $\phi=54\%$. Like Fig.~2b (main text) the centers of member tetrahedra are depicted and they are colored according to their environment: twelvefold ring (purple), capping PDs (green) and interstitials (blue). Compared to the quasicrystal shown in Fig. 2b (main text), the assembled quasicrystal has more defects and is yet to fully crystallize in the layers above and below the ones shown in the figure. This explains the abundance of rhombs and zippers that have shown to be important in the process of crystallization.}
\end{figure*}

\end{document}